\documentclass[reprint,aip,showpacs,apl,tightenlines,amsmath,amssymb,superscriptaddress]{revtex4-1}
\usepackage{graphicx}%
\usepackage{float}
\usepackage[T1]{fontenc}
\usepackage{color} 
\usepackage[colorlinks]{hyperref}
\usepackage{amsmath}
\usepackage{amssymb}
\usepackage{bm}
\usepackage[normalem]{ulem}

\begin{document}

\title{Low-frequency Raman scattering in WSe$_2$-MoSe$_2$ heterobilayers: Evidence for atomic reconstruction}	
\author{Johannes Holler}
\affiliation{Institut f\"ur Experimentelle und Angewandte Physik, Universit\"at Regensburg, D-93040 Regensburg, Germany}
\author{Sebastian Meier}
\affiliation{Institut f\"ur Experimentelle und Angewandte Physik, Universit\"at Regensburg, D-93040 Regensburg, Germany}
\author{Michael Kempf}
\affiliation{Institut f\"ur Experimentelle und Angewandte Physik, Universit\"at Regensburg, D-93040 Regensburg, Germany}
\affiliation{Institut f\"ur Physik, Universit\"at Rostock, D-18059 Rostock, Germany}
\author{Philipp Nagler}
\affiliation{Institut f\"ur Experimentelle und Angewandte Physik, Universit\"at Regensburg, D-93040 Regensburg, Germany}
\author{Kenji Watanabe}
\affiliation{Research Center for Functional Materials, National Institute for Materials Science, Tsukuba Ibaraki 305-0044, Japan}
\author{Takashi Taniguchi}
\affiliation{International Center for Materials Nanoarchitectonics, National Institute for Materials Science, Tsukuba Ibaraki 305-0044, Japan}
\author{Tobias Korn}
\affiliation{Institut f\"ur Physik, Universit\"at Rostock, D-18059 Rostock, Germany}
\author{Christian Sch\"uller}%
\email{christian.schueller@ur.de}
\affiliation{Institut f\"ur Experimentelle und Angewandte Physik, Universit\"at Regensburg, D-93040 Regensburg, Germany}

\date{\today}

\begin{abstract}
 We investigate WSe$_2$-MoSe$_2$ heterobilayers with different twist angles $\theta \pm \delta$ between the two layers, by low-frequency Raman scattering. In sufficiently aligned samples with $\theta=0^\circ$, or $\theta=60^\circ$, and $\delta \lesssim 3^\circ$, we observe an interlayer shear mode (ISM), which is a clear sign of a commensurate bilayer structure, i.e., the layers must undergo an atomic reconstruction to form R-type or H-type stacking orders. We find slightly different ISM energies of about 18~cm$^{-1}$ and 17~cm$^{-1}$ for H-type and R-type reconstructions, respectively, independent of the exact value of $\theta\pm \delta$. Our findings are corroborated by the fact that the ISM is not observed in samples with twist angles, which deviate by $\delta > 3^\circ$ from $0^\circ$ or $60^\circ$. This is expected, since in such incommensurate structures, with the possibility of Moir$\acute{\text{e}}$-lattice formation, there is no restoring force for an ISM. Furthermore, we observe the ISM even in sufficiently aligned heterobilayers, which are encapsulated in hexagonal Boron nitride. This is particularly relevant for the characterization of high-quality heterostructure devices.

\end{abstract}
\maketitle

The great appeal of van-der-Waals materials is the possibility to fabricate artificial multilayer structures, consisting of different materials, with arbitrary but well controlled relative crystal orientations \cite{Geim2013}. This offers new, and in some cases unexpected functionalities\cite{Cao2018}. Within the huge family of van-der-Waals materials, the semiconducting transition-metal dichalcogenides (TMDCs) have attracted tremendeous attention during the past decade\cite{Koperski2017,Wang2018}. In the monolayer form most of them represent direct-bandgap semiconductors\cite{Mak2010} with huge exciton binding energies\cite{Chernikov2014}, oscillator strengths\cite{Poellmann2015}, and spin-valley locking\cite{Xiao2012,Mak2012}. In recent years, heterobilayer structures with staggered type-II band-edge alignment \cite{Ozcelik2016,Kosmider2013,Kang2013} have attracted considerable interest, since in those structures interlayer excitons can form \cite{Fang2014,Rivera2015,Rivera2016,Kunstmann2018} due to fast charge separation of optically excited electron-hole pairs into the two constituent layers. For momentum-allowed interlayer transitions, the two constituent layers have to be crystallographically aligned, either in the H-type stacking configuration, where the layers are rotated by $\theta=60^\circ$ relative to each other, or in R-type stacking with $\theta=0^\circ$. Very recently, a strong focus in this research area has been on the exploration of possible  Moir$\acute{\text{e}}$-superlattice effects on interlayer excitons in heterobilayer structures\cite{Zhang2017,Zande2014,Alexeev2019,Jin2019,Seyler2019,Tran2019}. Due to the different lattice constants of the constituent materials in heterobilayers, Moir$\acute{\text{e}}$ structures are expected to form even for perfectly aligned structures, if rigid lattices of the constituent layers are assumed\cite{Kumar2015}. Moreover, the Moir$\acute{\text{e}}$-lattice period would decrease very quickly with increasing twist-angle deviation $\delta$ from $\theta =0^\circ$ or $60^\circ$, and would be smaller than the diameter of an interlayer exciton in, e.g., a WSe$_2$-MoSe$_2$ heterobilayer, of typically 3-4 nm, for $\delta \gtrsim 5^\circ$.

\begin{figure}[t!]
	\includegraphics[width= 0.45\textwidth]{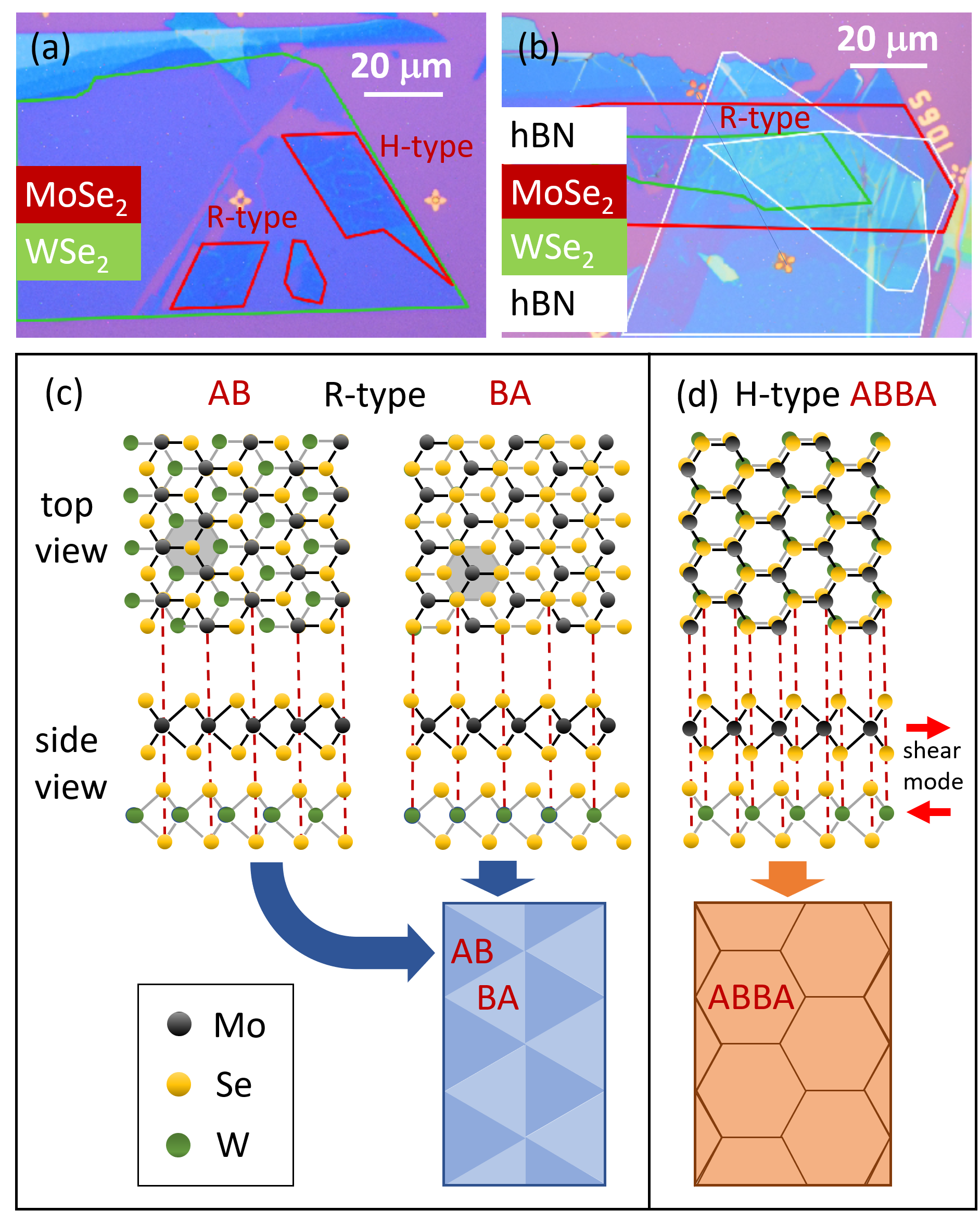}
	\caption{
		(a) Microscope image of an investigated heterobilayer sample. The red and green lines mark the MoSe$_2$ and WSe$_2$ monolayer regions, respectively. (b) Microscope image of a heterobilayer, encapsulated in hBN layers, which are oulined by white lines.
		(c) Left: Schematic of the atomic arrangement in an AB reconstructed R-type heterobilayer, where the metal atom (A) in the upper layer is above the chalcogens (B) in the lower layer. Right: Same for the energetically equivalent BA reconstructed R-type structure, where the chalcogens (B) in the upper layer are above the metal atoms (A) in the lower layer. The gray-shaded hexagon marks the same unit cell of the lower layer, while the upper layer is fixed. The lower panel shows the expected triangular domain pattern with AB- and BA R-type domains, as reported in Ref.~\onlinecite{Rosenberger20}.		
		(d) Atomic arrangement of a reconstructed ABBA H-type heterobilayer, where the upper layer is rotated by $60^\circ$ with respect to the lower one. The movement of the layers in the interlayer shear mode is indicated by red arrows. The lower panel shows a schematic picture of the expected hexagonal domain structure with ABBA atomic reconstruction\cite{Rosenberger20}.}
	\label{Fig1}
\end{figure}

Intriguingly, very recently it has been shown via conductive atomic force microscopy \cite{Rosenberger20} and transmission electron microscopy \cite{Weston2020} that in TMDC heterobilayers \cite{Rosenberger20,Weston2020} and homobilayers \cite{Weston2020} atomic reconstruction takes place for deviations $\delta \le 1^\circ$ (as reported in Ref.~\onlinecite{Rosenberger20}), or $\delta < 3^\circ$ (cf.~Ref.~\onlinecite{Weston2020}) from $\theta=0^\circ$ or $\theta=60^\circ$. We note that similar reconstructions are also reported for bilayer graphene\cite{Sushko2019}. Assuming rigid lattices of the constituent layers, Moir$\acute{\text{e}}$ superlattices would form in these cases. However, in Refs.~\onlinecite{Rosenberger20,Weston2020} it was found that the bilayers reconstruct in domains with H-type or R-type stacking configurations, i.e., in the commensurate lattice configurations of perfectly aligned homobilayers. The domain formation was theoretically predicted for the first time in Ref.~\onlinecite{Carr2018} via conformational considerations, and indeed, density functional theory calculations confirmed that in the above cases the two stacking configurations are the energetically favorable lattice arrangements \cite{Rosenberger20,Weston2020,Wozniak2020}. For R-type stacking, two energetically degenerate commensurate configurations are possible for a heterobilayer (see Fig.~\ref{Fig1}c), whereas the H-type stacking has only one energetically favorable lattice arrangement (displayed in Fig.~\ref{Fig1}d). In the recently published experimental and theoretical works \cite{Rosenberger20,Weston2020}, it is reported that for R-type reconstruction, triangular domains form with the two degenerate AB and BA lattice configurations (see Fig.~\ref{Fig1}c for illustration). For the H-type atomic reconstruction, the domains have hexagonal shape, as illustrated in Fig.~\ref{Fig1}d. The size of the domains depends on the deviation angle $\delta$, it is in the range of several tens to one hundred nanometers \cite{Rosenberger20}.

Raman spectroscopy is an important noninvasive tool in materials science. It has been very successfully applied to graphene\cite{Ferrari} and many other single- and multi-layered van-der-Waals materials. For an overview of Raman experiments on TMDCs, see, e.g., the review articles Refs.~\onlinecite{Zhang2015,Liang2017,Tan}, and references therein. The first observation of interlayer shear modes (ISM) in TMDC multilayers was reported by Plechinger et al.~\cite{Plechinger12} for MoS$_2$. Most importantly, for the existence of an ISM, a restoring force for the rigid layer displacement is required. Therefore, so far bilayer ISM have only been observed in homobilayers with R-type or H-type stacking, since there, a restoring force for the ISM is present \cite{Lui2015,Lin2018,Puretzky2016} due to the atomic registry of the commensurate equal lattices. For twist angles $\theta$ other than $0^\circ$ or $60^\circ$ \cite{Lin2018,Puretzky2016}, or, for heterobilayers \cite{Lui2015} and hetero-multilayers \cite{Lin2019}, where there is no restoring force for an ISM expected, only interlayer breathing modes (IBM) are reported so far for room-temperature experiments\cite{Lui2015,Lin2018,Puretzky2016,Lin2019}. For breathing modes, the van-der-Waals force between the layers plays the role of the restoring force.

In this work we employ low-frequency Raman scattering (LFRS) for the investigation of WSe$_2$-MoSe$_2$ heterobilayers with different twist angles. In sufficently aligned heterobilayers, we observe an ISM, which is clear evidence for a commensurate lattice arrangement, as provided by R-type and by H-type stackings. Very interestingly, we find slightly different ISM energies for H-type and R-type reconstructions, which offers the perspective of optical identification of these atomic reconstructions via contactless LFRS experiments. Furthermore, we observe the ISM even in sufficiently aligned heterobilayers, which are encapsulated in hexagonal Boron nitride (hBN). These results show the potential to identify commensurate stacking configurations even in buried heterostructures by noninvasive, contactless LFRS.

The Raman experiments are performed in an optical scanning-microscope setup at room temperature under ambient conditions. For excitation, a 532 nm laser line with 2.5 mW power is used. The laser is focused to a spot of $\sim 1$~$\mu$m diameter by a 100x microscope objective. For stray-light reduction, we use a Bragg-filter set. The spectra are analyzed with a grating spectrometer, and a Peltier-cooled CCD camera is used for detection.

The heterostructure samples are prepared on silicon wafers with SiO$_2$ layer via mechanical exfoliation, using Nitto tape, and a deterministic all-dry transfer technique \cite{Castellanos}, employing polydimethylsiloxane (PDMS) stamps. 
Figure \ref{Fig1}a shows a microscope image of one of the bare heterobilayer samples, investigated in this work. 
In Fig.~\ref{Fig1}b, an image of one of two hBN-encapsulated heterobilayers, investigated in this work, is displayed. 

For identification of $0^\circ$ and $60^\circ$ alignment in the first place, we use photoluminescence (PL) spectroscopy of the interlayer excitons in high magnetic fields\cite{unpublished} (cf.~Ref.~\onlinecite{Seyler2019}). As will be shown below, an important result of the present work is that the orientation ($0^\circ$ or $60^\circ$) can be determined by the energy of the ISM, if atomic reconstruction has taken place. We note that these results coincide perfectly with our assignments from magneto PL.

 \begin{figure}[t!]
	\includegraphics[width=0.45\textwidth]{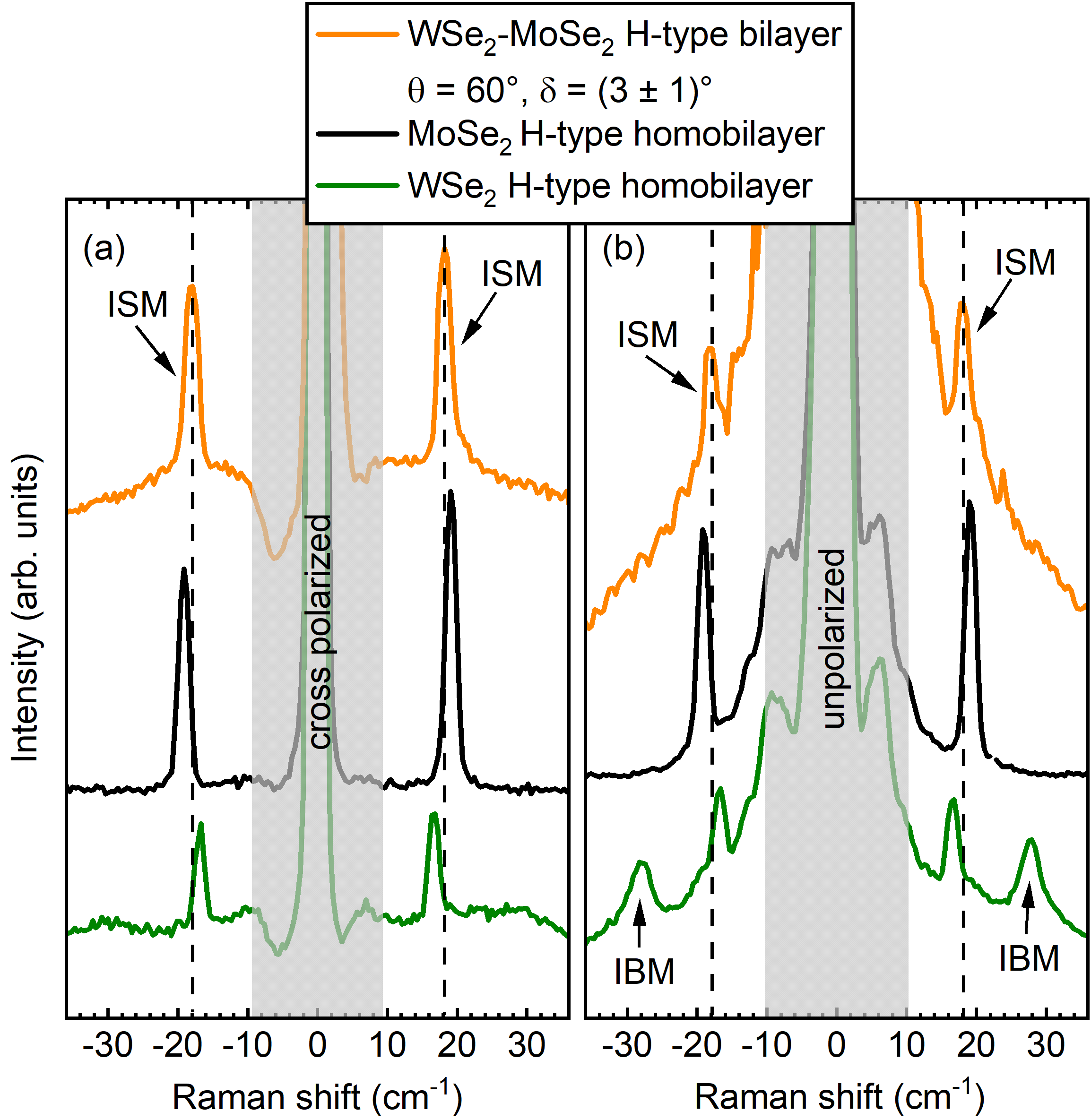}
	\caption{
		(a) Cross-polarized low-frequency Stokes- and Antistokes Raman spectra of exfoliated natural (H-type) WSe$_2$- and MoSe$_2$ homobilayers (green and black spectrum, respectively), and of a WSe$_2$-MoSe$_2$ heterobilayer with $\theta = 60^\circ$ and $\delta \sim 3^\circ$ (orange spectrum). (b) Same as (a) but for unpolarized detection.}
	\label{Fig2}
\end{figure}

We begin our discussion by comparing spectra of two exfoliated homobilayers to a heterobilayer. Figure \ref{Fig2}a shows linearly cross-polarized Stokes and Antistokes Raman spectra of WSe$_2$- and MoSe$_2$ homobilayers (green and black spectra), and of a WSe$_2$-MoSe$_2$ heterobilayer with $\theta = 60^\circ$ and $\delta \sim 3^\circ$. The spectra are not normalized but shifted vertically for better comparison. The two H-type homobilayer samples show the well-known ISM \cite{Chen2015} at energies of $\sim 17.7$ cm$^{-1}$ and $\sim 19.2$ cm$^{-1}$ for WSe$_2$ and MoSe$_2$, respectively. We use crossed linear polarizations, since the ISM is allowed in both configurations, parallel and crossed linear polarizations\cite{Plechinger12,Chen2015}, and the laser-stray-light reduction is much better in the latter case. Surprisingly, also the heterobilayer shows a strong Raman peak, which is energetically in between the peaks of the two homobilayers (indicated by a vertical dashed line in Fig.~\ref{Fig2}a). Therefore, it is highly likely that this Raman peak at an energy of $(18.0\pm 0.1)$ cm$^{-1}$ is an ISM of the heterobilayer. So far it was argued in literature \cite{Lui2015,Lin2018} that for heterobilayers with incommensurate lattice arrangements an ISM is not possible because of the lack of a restoring force for the rigid layer displacement. However, as mentioned above, recent reports show\cite{Rosenberger20,Weston2020} that both lattices can undergo atomic reconstruction for energetic reasons, and commensurability is restored. Therefore, we interpret the Raman peak at 18 cm$^{-1}$ in the orange spectrum in Fig.~\ref{Fig2}a as the ISM of the WSe$_2$-MoSe$_2$ heterobilayer. Consequently, this measurement provides evidence for the H-type atomic reconstruction. For comparison, we plot in Fig.~\ref{Fig2}b Raman spectra of the same samples for unpolarized detection of the inelastically scattered light. In these spectra, the ISM peaks of all samples are reproduced, as expected from the selection rules. In addition, the IBM is visible at an energy of about 27.7 cm$^{-1}$ in the spectrum of the WSe$_2$ homobilayer (green spectrum in Fig.~\ref{Fig2}b). For the other two samples, the scattering probability for the IBM at the used laser energy is obviously too small to be observed in the measurement. One can also clearly recognize the much stronger laser stray-light signal in the unpolarized Raman spectra in Fig.~\ref{Fig2}b. The IBM should in principle also exist for the heterobilayer, since for the IBM the van-der-Waals force between the layers plays the role of a restoring force for the rigid layer breathing oscillation\cite{Lui2015}. Since our focus in this work is the ISM, which requires commensurate lattices, and for better stray-light suppression, we stay in the following with crossed-linear polarization configurations. We note that we have measured several spots (typically 3) on all investigated samples. We find very similar spectra with exactly the same ISM energy on all investigated spots of a given sample, without any exception. The spectra shown in the plots are thus representative.

\begin{figure}[t!]
	\includegraphics[width= 0.45\textwidth]{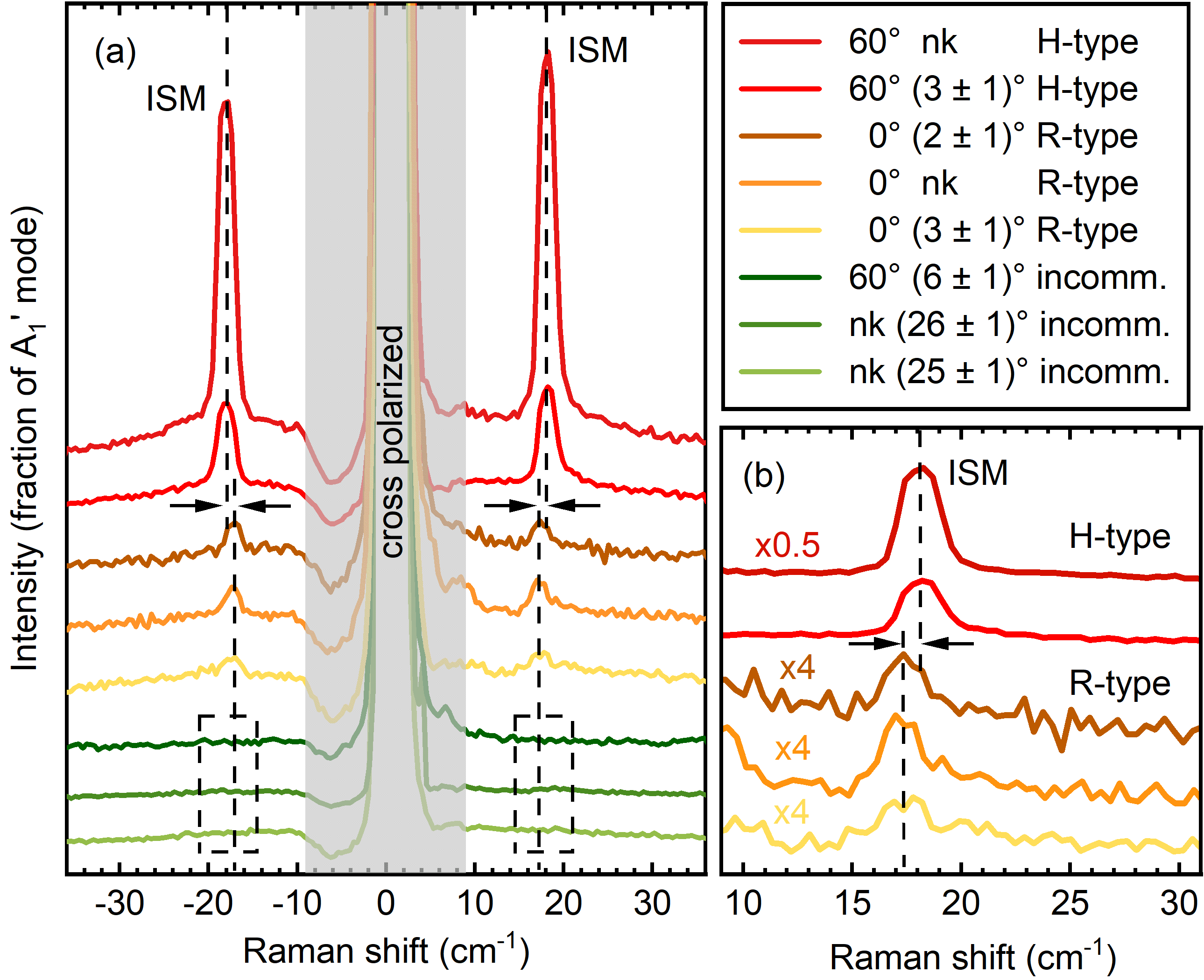}
	\caption{(a) Representative cross-polarized low-frequency Raman spectra of eight different WSe$_2$-MoSe$_2$ heterobilayers. The respective relative twist angles $\theta$ of the heterobilayers are given in the legend on the right-hand side. The measured deviations $\delta$ are given, together with estimated errors, in brackets. 'nk' means 'not known'. (b) Blowup of the Stokes region for the samples with lattice reconstruction. Some of the spectra are renormalized as indicated.}
	\label{Fig3}
\end{figure}

For more detailed analysis we show in Fig.~\ref{Fig3}a representative Stokes-Antistokes spectra for all investigated heterobilayers without hBN encapsulation. For a quantitative comparison, all spectra have been normalized to the intensities of the A$_1$' monolayer optical phonons of the respective samples. The A$_1$' phonons are at much larger Raman shifts, which is not displayed in the low-frequency spectra in Fig.~\ref{Fig3}a. For a comparison of normalized Raman spectra also over the energy range of the optical phonons, see Fig.~S1 in the Supplemental Material. The legend of Fig.~\ref{Fig3} contains the twist angles $\theta$, and the deviations $\delta$ in brackets, if known. If the corresponding values are not known for a sample, this is indicated by 'nk'. Additionally, our interpretations of the lattice arrangements are also mentioned as H-type, R-type, or incommensurate in the legend. For most of the samples, the twist angles could be determined via second-harmonic generation microscopy\cite{SHG,Kunstmann2018}. The procedure how the angles are determined is described in more detail in the Supplemental Material. For an accurate determination, large enough monolayer parts of each material are required. This is not given for all samples. Therefore, some of the angles are not precisely known. For those samples we have to rely on the accuracy of parallel sample-edge alignment in the preparation processes, which is about $\pm 3^\circ$.

The important conclusions from Fig.~\ref{Fig3}a are the following: (i) An ISM is detectable for all samples with twist-angle deviation $\delta\lesssim 3^\circ$ from $\theta =0^\circ$ or $\theta = 60^\circ$ (yellow, two orange, and two red spectra in Fig.~\ref{Fig3}a). (ii) For $\theta=0^\circ$ (R-type reconstruction), the intensities of the ISM are much smaller, in our measurements by a factor of about 5 to 10, than for $\theta=60^\circ$ (H-type reconstruction). (iii) The energy of the ISM for R-type reconstruction is about ($17.4\pm 0.1$) cm$^{-1}$, while for H-type reconstruction it is ($18.0\pm 0.1$) cm$^{-1}$. (iv) For deviation angles $\delta > 3^\circ$ no ISM signal is detectable (green spectra in Fig.~\ref{Fig3}a), though the optical phonons of the constituent layers, which appear at higher Raman shifts, are measured with comparable strengths (see supplemental Fig.~S1). The findings (ii) and (iii) are consistent with the behavior of the ISM in MoSe$_2$ homobilayers, where, similarly, the intensity is much lower and the energy slightly smaller for R-type than for H-type stacking \cite{Puretzky2015}. The reason for this is the smaller interlayer bond polarizability for the R-type configuration\cite{Puretzky2015}. To emphasize (iii) -- the difference of the ISM energies for R-type and H-type reconstructions -- we show in Fig.~\ref{Fig3}b a blowup of the Stokes side of the corresponding spectra. It can be clearly seen that the energies of the ISM are constant for a given configuration (R-type or H-type), irrespective of the exact deviation angle $\delta\lesssim 3^\circ$. This confirms our interpretation that the observation of an ISM in a heterobilayer sample is evidence for an atomic reconstruction: If the reconstruction takes place, the ISM has a well-defined energy.
One could imagine that the intensities of the ISM in Fig.~\ref{Fig3}a are related to the deviation angle $\delta$, and, hence, to the size of the domains: The larger $\delta$, the smaller are the domains, the larger is the areal fraction of the domain walls, and the smaller is the area with atomic reconstruction, where the ISM is defined. To prove such a speculative relation, more experiments on much more samples with known twist angles are required in future work.

\begin{figure}[t!]
	\includegraphics[width= 0.35\textwidth]{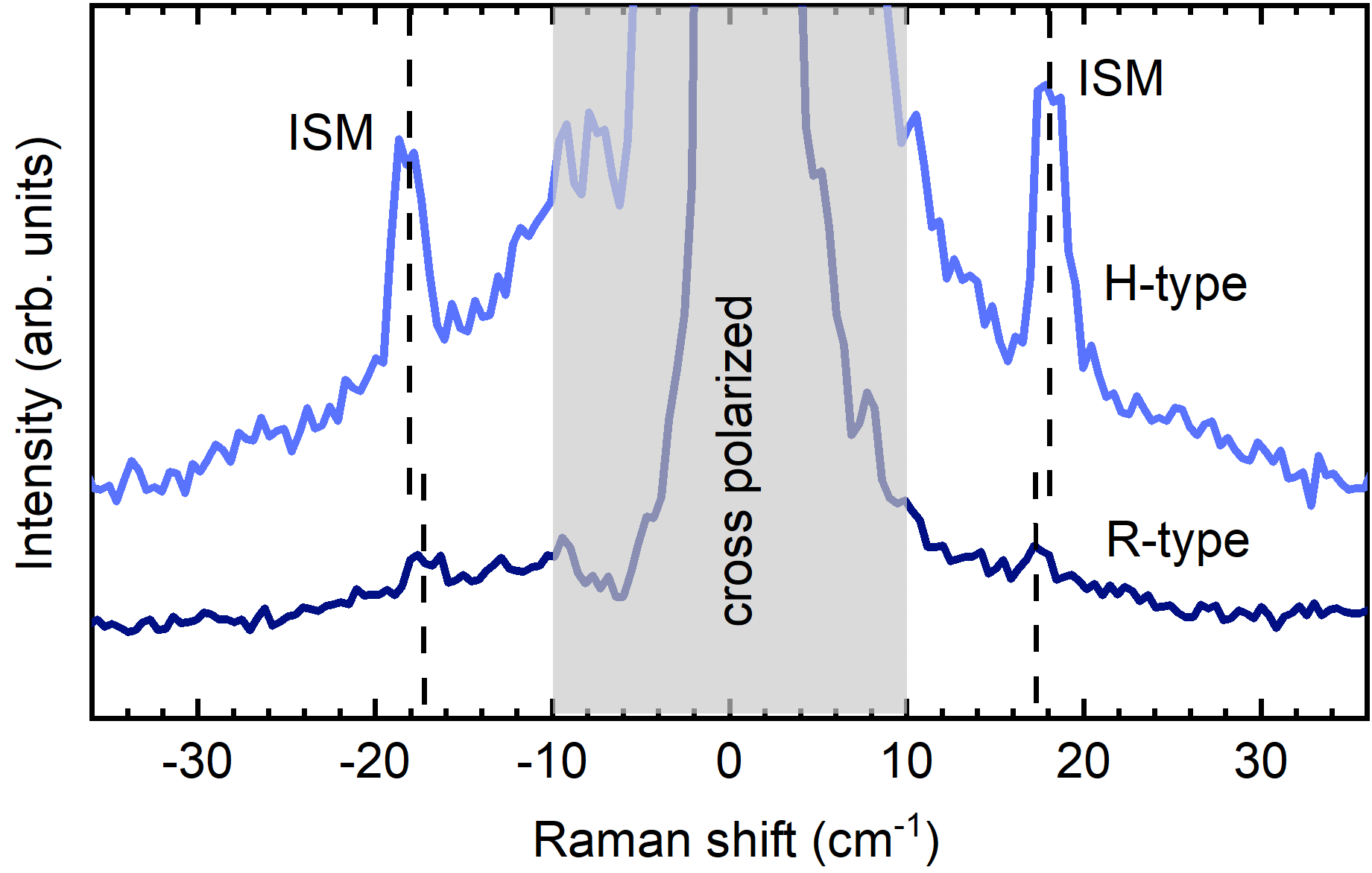}
	\caption{Representative Raman spectra of two WSe$_2$-MoSe$_2$ heterobilayer with $60^\circ+\delta $ (light blue spectrum), and $0^\circ+\delta$ (dark blue spectrum), which are encapsulated in hBN multilayers.}
	\label{Fig4}
\end{figure}

Finally, in Fig.~\ref{Fig4} Raman spectra of two heterobilayers, which are encapsulated in hBN multilayers, are displayed. Also in these samples an ISM can be observed. Moreover, for both reconstructions, R-type and H-type, the energies of the ISM are the same as for the bare heterobilayer samples. Obviously, the hBN lattice is completely incommensurate with the MoSe$_2$ and WSe$_2$ lattices. Therefore, there is no restoring force interaction between the encapsulating layers and the heterobilayer, and, hence, the ISM is not disturbed by the presence of the hBN layers. The experiments in Fig.~\ref{Fig4} clearly demonstrate this effect. This is a very important further result of our investigation: Even atomic reconstruction in buried heterobilayers can be detected and identified by noninvasive and contactless LFRS.  

In conclusion, we have demonstrated that atomic reconstruction in TMDC heterobilayers can be detected by low-frequency Raman spectroscopy, by the presence of an ISM. Even the type of reconstruction - R-type or H-type - can be identified via the energy of the ISM. An important further finding is that hBN encapsulation of the heterobilayer does not significantly influence the proposed method. The latter is a very important information for the design of technologically relevant, high-quality heterostructures devices.

\begin{acknowledgments}
We greatfully acknowledge funding by the Deutsche 
Forschungsgemeinschaft (DFG, German Research Foundation) - Project-ID 
314695032 - SFB 1277 (subprojects B05 and B06), and projects KO3612-3 and KO3612-4. K.W. and T.T. acknowledge support from the Elemental Strategy Initiative conducted by the MEXT, Japan, Grant Number JPMXP0112101001,  JSPS
KAKENHI Grant Numbers JP20H00354 and the CREST(JPMJCR15F3), JST.
\end{acknowledgments}

\section*{Supplemental Material}
	Raman spectra over a broader energy range, the determination of the twist angles of the heterobilayers via second-harmonic-generation microscopy, and an overview over the investigated samples are provided in the Supplemental Material.\\
	
	The data that support the findings of this study are available from the corresponding author upon reasonable request.

\end{document}


\title{Supplemental Material for:\\
	Low-frequency Raman scattering in WSe$_2$-MoSe$_2$ heterobilayers: Evidence for atomic reconstruction}	
\author{Johannes Holler}
\author{Sebastian Meier}
\author{Michael Kempf}
\author{Philipp Nagler}
\affiliation{Institut f\"ur Experimentelle und Angewandte Physik, Universit\"at Regensburg, D-93040 Regensburg, Germany}
\author{Kenji Watanabe}
\affiliation{Research Center for Functional Materials, National Institute for Materials Science, Tsukuba Ibaraki 305-0044, Japan}
\author{Takashi Taniguchi}
\affiliation{International Center for Materials Nanoarchitectonics, National Institute for Materials Science, Tsukuba Ibaraki 305-0044, Japan}
\author{Tobias Korn}
\affiliation{Institut f\"ur Physik, Universit\"at Rostock, D-18059 Rostock, Germany}
\author{Christian Sch\"uller}%
\email{christian.schueller@ur.de}
\affiliation{Institut f\"ur Experimentelle und Angewandte Physik, Universit\"at Regensburg, D-93040 Regensburg, Germany}

\date{\today}

\maketitle

\begin{figure}[t!]
	\includegraphics[width= 0.45\textwidth]{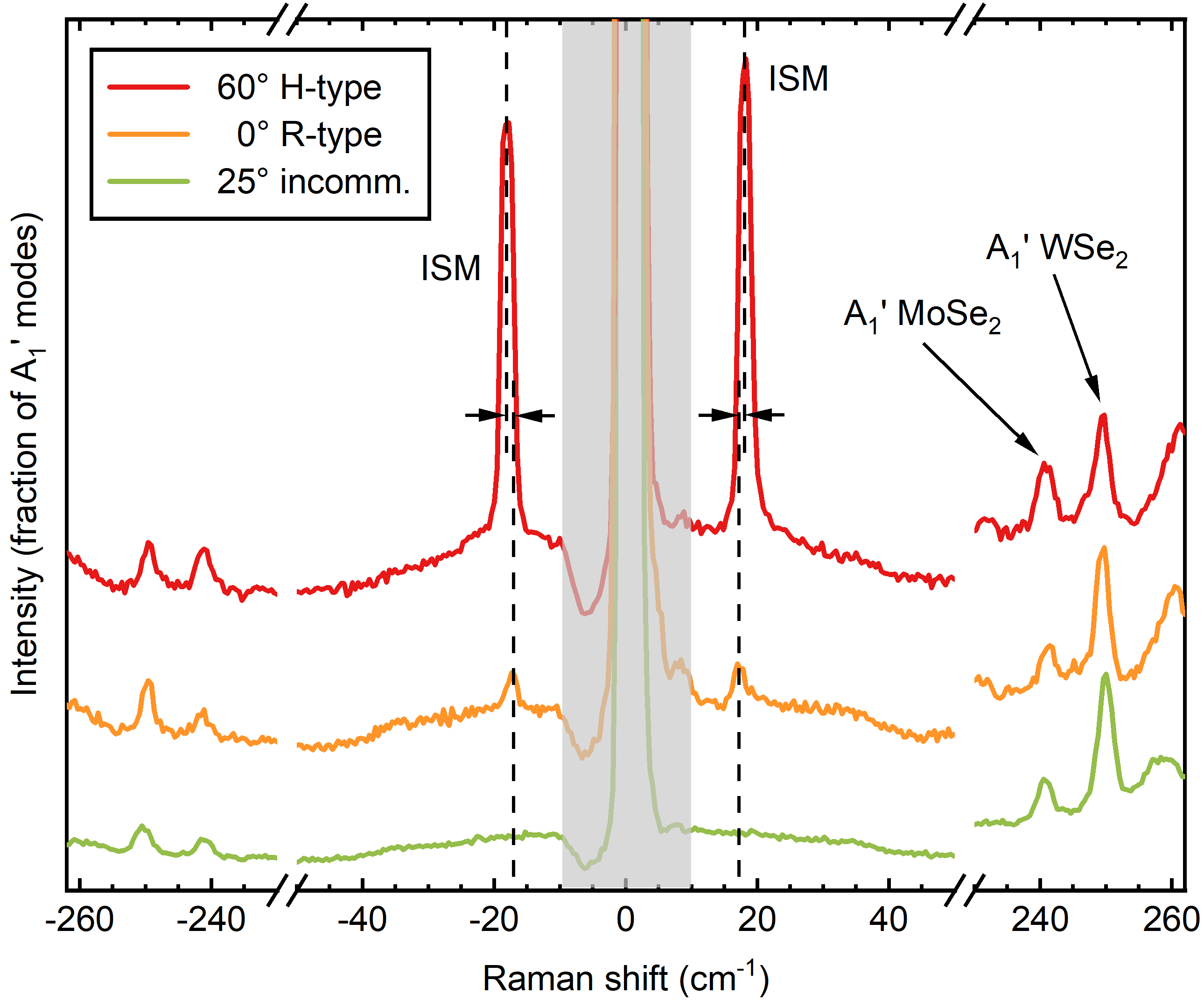}
	\caption{Linear cross-polarized Stokes- and Antistokes Raman spectra of 3 selected WSe$_2$-MoSe$_2$ heterobilayer samples for the energy range of the ISM and the A$_1$' optical phonons of the constituent layers. The color code is the same as for Fig.~3 of the manuscript.
	}
	\label{FigS1}
\end{figure}

In the manuscript, low-fequency Raman spectra are shown for the frequency range of the interlayer shear modes (ISM), only. In Figure \ref{FigS1} we show selected spectra for the three discussed cases (H-type and R-type reconstructions, incommensurate lattices) for the low-frequency range and also for the range of the optical A$_1$' phonons of the constituent materials (the color code is the same as in Fig.~3 of the manuscript). For quantitative comparison of the phonon intensities, the spectra are normalized to the sum of the intensities of the A$_1$' phonons of the WSe$_2$ and the MoSe$_2$ layer. The same is done with all the spectra, displayed in Fig.~3 of the manuscript.

\begin{figure*}[t!]
	\includegraphics[width= 0.65\textwidth]{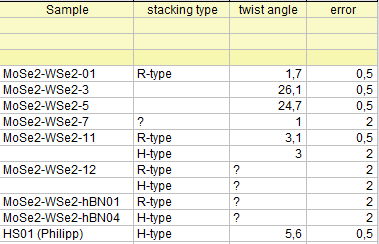}
	\caption{Microscope image of one of the investigated samples. The polar plot in the right panel shows the angular dependence of the measured SHG signals. The spots on the monolayers, where the measurements were taken, are indicated by arrows.}
	\label{FigS2}
\end{figure*}

The twist angle is determined by polarized second harmonic generation (SHG) measurements\cite{SHG} on the respective monolayers of the heterostructures. Our SHG setup is described in more detail in the Supplemental Material of Ref.~\onlinecite{Kunstmann2018}. Figure \ref{FigS2} shows an example for the MoSe$_2$-WSe$_2$ heterobilayer with deviation angle $\delta = 25^\circ$. The two spots on the monolayer parts, where the SHG measurements are taken, are marked by arrows. The polar plot presents a section of the SHG measurement, the complete measurement was done over an angle of 720$^\circ$ and fitted via Origin software. The maxima of the SHG measurement correspond to the crystallographic orientation of the monolayers and therefore the twist angle can be determined\cite{SHG}. The estimated error is 1$^\circ$. Not all samples have monolayer regions of both materials, hence, for some samples the twist angle was estimated by the optical microscope pictures, where the crystallographic orientation can be approximated by the edges of the sample. The error of this method depends on the quality of the sample edge and the optical picture. Note that with these methods we cannot distinguish between R-type and H-type stackings.

In Table I, all investigated samples are listed. The twist angles $\theta$, as well as the deviation angles $\delta$ are given. If a value could not be accurately determined, this is indicated by 'nk' for 'not known'. By check marks it is also indicated on which samples SHG measurements were performed, whether or not the twist angles were also determined by the optical microscope images, which samples were annealed in vacuum at a temperature of $100^\circ$ C for at least one hour, and which are encapsulated in hexagonal Boron nitride layers.

\begin{table*}[]
	\begin{tabular}{|l|c|c|c|c|c|c|}
		\hline 
		Sample name & $\theta$ ($^\circ$) & $\delta$ ($^\circ$) & SHG & optical & annealed & hBN \\ 
		\hline 
		HS01 & 60 & 6$\pm$1 & \checkmark & \checkmark & \checkmark &  \\ 
		\hline 
		MoSe$_2$-WSe$_2$-01 & 0 & 2$\pm$1 & \checkmark & \checkmark &  &  \\ 
		\hline 
		MoSe$_2$-WSe$_2$-03 & nk & 26$\pm$1 & \checkmark &  & \checkmark &  \\ 
		\hline 
		MoSe$_2$-WSe$_2$-05 & nk & 25$\pm$1 & \checkmark &  & \checkmark &  \\ 
		\hline 
		MoSe$_2$-WSe$_2$-11 region 1 & 0 & 3$\pm$1 & \checkmark & \checkmark &  &  \\ 
		\hline 
		MoSe$_2$-WSe$_2$-11 region 2 & 60 & 3$\pm$1 &  & \checkmark &  &  \\ 
		\hline 
		MoSe$_2$-WSe$_2$-12 region 1 & 0 & nk &  &  &  &  \\ 
		\hline 
		MoSe$_2$-WSe$_2$-12 region 2 & 60 & nk &  &  &  &  \\ 
		\hline 
		MoSe$_2$-WSe$_2$-hBN01 & 0 & nk &  &  &  & \checkmark \\ 
		\hline 
		MoSe$_2$-WSe$_2$-hBN04 & 60 & nk &  &  &  & \checkmark \\ 
		\hline 
	\end{tabular}
	\caption{List of samples used in this work with twist angle $\theta\pm\delta$, determined via SHG and/or optical microscopy. It is also shown which samples were annealed and which are encapsulated in hBN. 'nk' means 'not known'.}
	\label{table1}
\end{table*}